\documentclass[aps,prl,twocolumn,showpacs,groupedaddress,superscriptaddress,longbibliography]{revtex4}

\usepackage{amsmath}
\usepackage{color}
\usepackage{amsfonts}
\usepackage{epsf}
\usepackage{graphicx}
\usepackage{ulem}
\definecolor{red}{rgb}{1,0,0}
\definecolor{blue}{rgb}{0,0,1}

\usepackage{mathrsfs}
\begin{document}

\title{Actomyosin contraction induces droplet motility}

\author{T. Le Goff}
\email{thomas.le-goff@ed.ac.uk}
\affiliation{SUPA, School of Physics and Astronomy, University of Edinburgh, Peter Guthrie Tait Road, Edinburgh, EH9 3FD, UK}
\author{B. Liebchen}
\email{liebchen@hhu.de}
\affiliation{SUPA, School of Physics and Astronomy, University of Edinburgh, Peter Guthrie Tait Road, Edinburgh, EH9 3FD, UK}
\affiliation{Institute for Theoretical Physics II: Soft Matter, Heinrich-Heine University D\"usseldorf, Universit\"atsstrasse 1, 40225, D\"usseldorf, Germany} 
\author{D. Marenduzzo}
\email{dmarendu@ph.ed.ac.uk}
\affiliation{SUPA, School of Physics and Astronomy, University of Edinburgh, Peter Guthrie Tait Road, Edinburgh, EH9 3FD, UK}

\begin{abstract}
While cell crawling on a solid surface is relatively well understood, and relies on substrate adhesion, some cells can also swim in the bulk, through mechanisms that are still largely unclear. Here, we propose a minimal model for in-bulk self-motility of a droplet containing an isotropic and compressible contractile gel, representing a cell extract containing a disordered actomyosin network. 
In our model, contraction mediates a feedback loop between myosin-induced flow and advection-induced myosin accumulation, which leads to clustering and a locally enhanced flow. 
Interactions of the emerging clusters with the droplet membrane break flow symmetry and set the whole droplet into motion. Depending mainly on the balance between contraction and diffusion, this motion can be either straight or circular.
Our simulations and analytical results provide a framework allowing to study in-bulk myosin-driven cell motility in living cells and to design synthetic motile active matter droplets.
\end{abstract}

\maketitle


{\it Introduction} -- Understanding the rules governing cell motion is a fascinating problem in biophysics, because the engine governing motility is purely self-organised~\cite{bray,phillips}. 
The mechanism of cell motility is also of major biomedical relevance, as this process is central to the self-assembly of tissues in a growing embryo, is required for wound healing, and is important to understand the pathway through which cancer cells metastatise.

Crawling on a solid substrate~\cite{bray,schaub,phillips,stricker,crawling,goff2016} is the motility mode currently best characterised, both experimentally and theoretically. It requires polymerisation of the actin cytoskeletal network, which pushes the cell forward by ratcheting the motion of its plasma membrane~\cite{brownianratchet}. 
For this mechanism to work, the actin cytoskeleton needs to be anchored to the substrate at least transiently, to avoid back-slip of the whole network following polymerisation. 
Indeed, anchoring points are well documented for crawling cells: these are ``focal adhesions'', formed by clusters of transmembrane proteins binding to the substrate~\cite{phillips,crawling}. 
This mechanistic understanding of cell crawling has been described in various successful models, quantitatively explaining, in particular, the locomotion of keratocyte cells on a substrate~\cite{kruse2006,mogilner}.

Crucially, however, some cells can even move through tissue or the extracellular matrix~\cite{3Dcellmotility}, where we have no underlying substrate. 
Cancer cells invading a 3D matrigel have a spherical morphology, possess no lamellipodium, and show an accumulation of actin at their back~\cite{keller,piel} rather than the front, making it unlikely that actin polymerisation is directly responsible for cell locomotion. 
This example suggests that the mechanism for in-bulk cell motion, which is not understood in detail~\cite{3Dcellmotility}, must be fundamentally different from that of crawling in 2D. 
The challenge of moving in absence of a substrate can be appreciated by comparing the mechanism allowing birds to fly through 3D space, with that exploited by animals to walk on the ground.

Our goal in this work is to provide a model and mechanism for cell motility in bulk, which is both minimal and generic. 
Since myosin is currently the best candidate to provide the engine for 3D cell motility, through some form of ATP-dependent contractility~\cite{elsen,recho}, we model {\it isotropic} contraction of an actomyosin gel confined in a droplet, mimicking a cell or a cell extract (i.e., a bag of actomyosin enclosed by a membrane, without regulatory proteins). 
Previous work proposed related models of contractility-induced motility~\cite{elsen,carl,carl2,recho,nickaeen2017,notenickaeen}: these studies, however, either invoked the rectification of splay fluctuations in droplets 
of {\it anisotropic} active nematic gels~\cite{elsen,carl,carl2}, or considered additional ingredients 
besides an actomyosin droplet, such as a thin cortex to which motors can bind dynamically~\cite{carl2}, or a frictional substrate~\cite{recho,nickaeen2017}.
In our minimal model, the mechanism for in-bulk motility is provided by contractility, favouring myosin clustering, and steric interaction with the enclosing membrane, breaking the symmetry of the density field spatially. 
 We find two modes of in-bulk motility, associated with either linear, or circular motion. 



{\it Model} -- 
To model in-bulk cell motility we describe a subcellular actomyosin gel as an isotropic active gel with a stress tensor,
\begin{equation}\label{stress}
\bar{\bar{\sigma}}=\mu\left[\nabla{\bf v}+(\nabla{\bf v})^{T}\right]+\left[\lambda(\nabla.{\bf v})+\mathcal{X} f(m)\right]\mathbb{I},
\end{equation}
where ${\bf v}$ is the velocity field of the active gel, $\mu$ is its dynamic viscosity, $\lambda$ the bulk viscosity and $\mathcal{X}$ measures myosin-induced contraction. The strength of contractility depends on the concentration of myosin motors, $m$, as 
$f(m)=\frac{m}{1+m}$,
ensuring saturation at large $m$~\cite{moore}. As actomyosin is contractile, 
${\mathcal{X}}>0$. [Note Eq.~\ref{stress} disregards passive contributions, which we assume negligible with respect to active ones as in~\cite{recho}.]
 
To model myosin transport, we use an advection-diffusion equation. Here, the local advection velocity of myosin may be different from that of the active gel, since motors can dynamically attach and detach from actin filaments with rate depending on the environment~\cite{adelstein}. 
We therefore introduce the dimensionless paramater $\alpha_m \in [0,1]$ to quantify the affinity of myosin with actin, where $\alpha_m=1$ means all motors are permanently attached to the actomyosin gel. Additionally, force balance (where inertial terms can be neglected at cellular scales) yields the following equations of motion for the myosin density field $m({\bf x},t)$ and the actomyosin velocity field ${\bf v}({\bf x},t)$:
\begin{equation}
\left\{
\begin{array}{l}
  \partial_tm=-\alpha_{m}{\bf \nabla}.(m{\bf v})+D_m{\bf \nabla}^2m,\\
  \gamma v_x=(2\mu+\lambda)\partial_x^2v_x+\mu\partial_y^2v_x+(\mu+\lambda)\partial_x\partial_yv_y\\
  \hspace{0.95cm}+\mathcal{X}\partial_xf(m),\\
  \gamma v_y=(2\mu+\lambda)\partial_y^2v_y+\mu\partial_x^2v_y+(\mu+\lambda)\partial_x\partial_yv_x\\
  \hspace{0.95cm}+\mathcal{X}\partial_yf(m),
\end{array}
\right.
\end{equation}
where $D_m$ is the myosin diffusion coefficient and $\gamma$ the friction coefficient which is $\ne 0$ only with an underlying substrate. 
We formulated our model in 2D to allow for systematic parameter sweeps -- 
extension to 3D is straightforward and should lead to analogous results. 

To reduce the parameter space to its essential dimensions, we now use $t_u=\mu/\mathcal{X}_0$ and $x_u=\sqrt{D_m \mu/\mathcal{X}_0}$ as time and space units where $\mathcal{X}_0$ is a reference value for contractility. We also introduce dimensionless parameters $\eta\equiv \lambda/\mu$ (ratio of bulk and dynamic viscosity), $\chi\equiv \mathcal{X}/\mathcal{X}_0$ (contraction strength), $\Gamma\equiv D_m \gamma/\mathcal{X}_0$ (reduced substrate friction, which vanishes without a substrate), and use the dimensionless fields $\tilde m=m/m_0$ (where $m_0$ is the average actomyosin density which is conserved under the dynamics) and $\tilde {\bf v}=\sqrt{\frac{\mu}{D_m \mathcal{X}_0}}{\bf v}$.

Inspired by previous works~\cite{shao,dreher}, we use a phase-field approach to model enclosure of actomyosin within a membrane, to mimic a cell or cell extract. Thus, we define a phase field $\phi({\bf x},t)$ and a corresponding equation of motion featuring two fixed points representing locally uniform phases: $\phi \approx 1$, representing the interior of the cell, and $\phi \approx 0$, representing the space outside it. 
 
In dimensionless units and in presence of the phase field, our minimal model reads (omitting tildes):
\begin{equation}
\left\{
\begin{array}{l}
  \partial_tm=-\alpha_{m}{\bf \nabla}.(m{\bf v})+{\bf \nabla}^2m+\epsilon_m{\bf \nabla}^2(\frac{\delta\mathcal{E}}{\delta m}),\\
  \Gamma v_x=(2+\eta)\partial_x^2v_x+\partial_y^2v_x+(1+\eta)\partial_x\partial_yv_y\\
  \hspace{0.95cm}+\chi\partial_xf(m),\\
  \Gamma v_y=(2+\eta)\partial_y^2v_y+\partial_x^2v_y+(1+\eta)\partial_x\partial_yv_x\\
  \hspace{0.95cm}+\chi\partial_yf(m),\\
  \partial_t\phi=D_{\phi}{\bf \nabla}^2\phi-\Gamma_{\phi}U'(\phi)-{\bf v}.{\bf \nabla}\phi,
\end{array}
\right.
\label{phasefieldequations}
\end{equation}
Here, we have introduced $\mathcal{E}(m,\phi)=\iint~\mathrm{d}{\bf r}\{(m^2+\alpha)[(1-\phi)^2+\alpha]\}^{1/2}$ as an effective energy to constrain myosin within the cell boundaries. 
The diffusivity $D_{\phi}$ quantifies the ability of the cell to oppose deformation, hence we call it deformation resistance. Its effect is similar to surface tension, which would, however, enter the equations of motion in a formally different way~\cite{folch,biben}. 
The term $U'(\phi)=\phi(\phi-1)[\phi-\frac{1}{2}-\alpha_0(\frac{V}{V_{tar}}-1)]$, is the derivative of the double-well potential $U$ whose fixed points $\phi=0$ and $\phi=1$ describe the outside and inside of the cell extract. 
The droplet interface (cell boundary) has a characteristic width of $(8D_{\phi}/\Gamma_{\phi})^{1/2}$. The term $\alpha_0(\frac{V}{V_{tar}}-1)$ restores the instantaneous cell volume $V=\iint\mathrm{d}{\bf r}~\phi^2(3-2\phi)$ towards a characteristic target volume $V_{tar}$. Finally, $-{\bf v}.{\bf \nabla}\phi$ represents the advection of the actomyosin network. 

To get an intuition for the order of magnitude of our model parameters, we set experimentally relevant length, time and viscosity scales for cell extracts and actomyosin droplets as $x_u\sim 1~\mu m$, $t_u\sim 1$ s, and $\mu\sim 10$ Pa.s~\cite{wottawah,norstrom}. These give $D_m\sim 1\mu m^2s^{-1}$, and $\mathcal{X}_0\sim 10$ Pa -- the former value is close to the {\it in vivo} myosin diffusion coefficient, to gauge the latter we note a myosin concentration of $m_0\sim 1-10 \mu M$~\cite{janson,norstrom} and a force per motor of 10 $pN$~\cite{norstrom,finer,bendix} creates a contractility of $\mathcal{X} \sim 3-30 \mathcal{X}_0$ Pa (calculated assuming a myosin size $\sim 50$ nm~\cite{bray}). 

\begin{figure}
\centering
\includegraphics[width=6.cm]{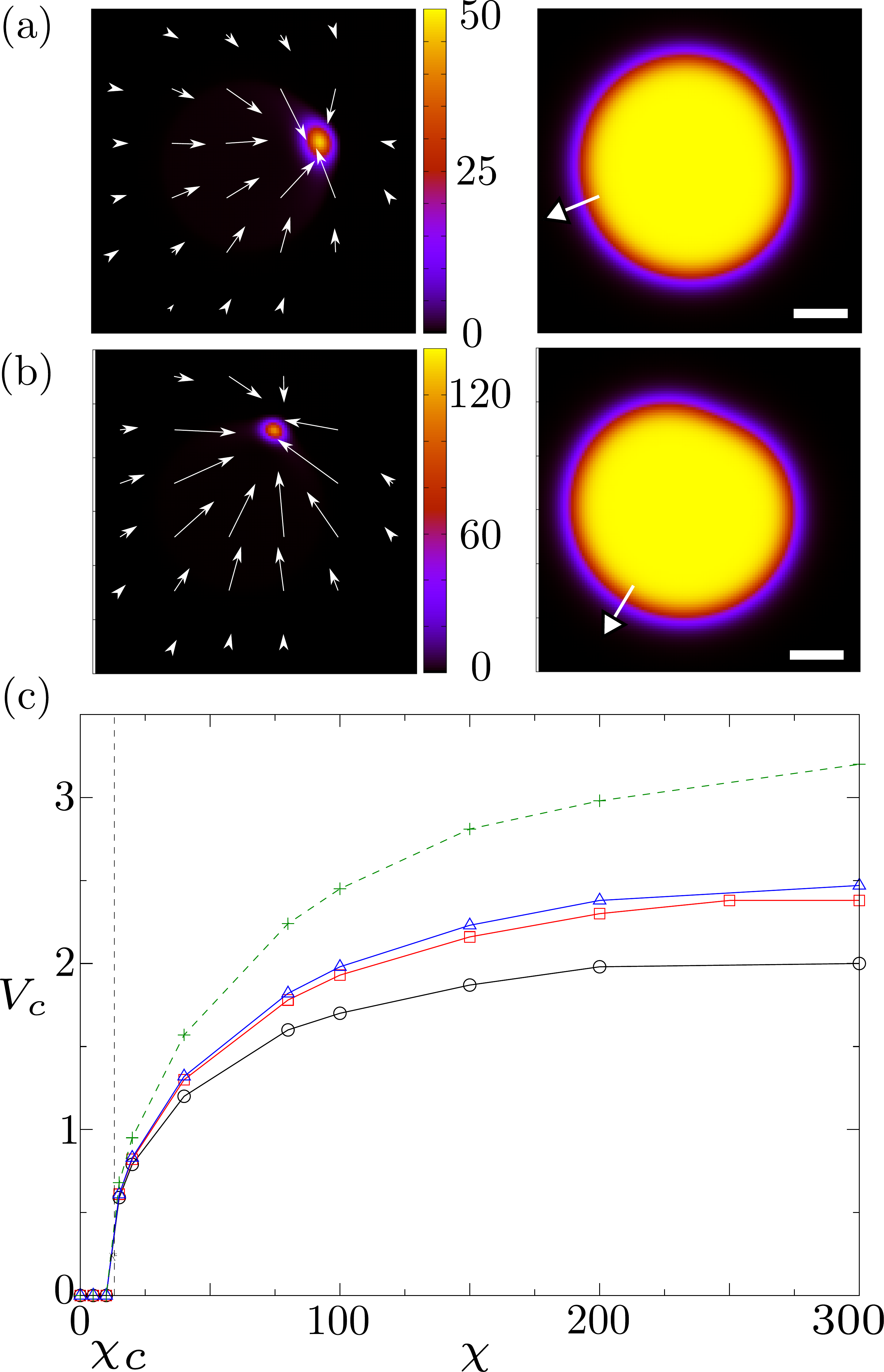}
\caption{(a,b) Left: concentration of myosin and ${\bf v}$-field for $V_{tar}=12.5$, $D_{\phi}=20$, $\alpha_{m}=1$ and $\gamma=0$. Right:  corresponding phase field. The white arrow gives the direction of cell motion and the scale bar is $1$. The contractility is $\chi=80$ in (a) and $\chi=200$ in (b). (c) Overall cell velocity as a function of $\chi$. The solid lines corresponds to simulations without solid friction and the dashed-line corresponds to simulations with $\gamma=0.2$. Black line and circles: $D_{\phi}=10$; red line and squares: $D_{\phi}=20$; blue line triangles: $D_{\phi}=25$; green dashed line and crosses: $D_{\phi}=25$.} 
\label{fig1}
\end{figure}


{\it Contractility induced cell-motility} --
To explore the dynamics of actomyosin droplets, we simulate Eqs.~(\ref{phasefieldequations}). As parameters, we use $\eta=-2/3$~\cite{batchelor}, $\epsilon_m=20$, $\alpha=10^{-4}$ and $\alpha_0=50$. We choose $D_{\phi}/\Gamma_{\phi}=160$, to fix the shape and width of the cell boundary throughout our simulations. 
%
We also set $\Gamma=0$ to study in-bulk motility, $m_0=1$, and choose initial conditions as $m=m_0+\delta m$ and ${\bf v}=\delta{\bf v}$ where $\delta m$ and $\delta{\bf v}$ represent small fluctuations.

We first consider the limit where myosin has a strong affinity with actin ($\alpha_m=1$). For small contractility $\chi$, the myosin remains uniform within the cell, which is stationary. However, when the contractility surpasses a threshold, a myosin spot forms at one edge of the cell (Figs.~1a,b and Suppl. Movie 1~\cite{SM}). 
While this spot grows, the cell deforms. Strikingly, it then starts to move away from the myosin spot, which now sits at its rear. Soon, the cell reaches a constant velocity and moves along a straight line (Suppl. Movie 1~\cite{SM}). 

Although we do not directly model the underlying solvent flow, a viable pattern is one in which it opposes actomyosin flow, to ensure that the whole system (actomyosin plus solvent) is incompressible. 
This is a realistic flow pattern as, at the timescales we consider, membranes can be regarded as permeable, hence solvent can move in and out of a cell extract freely~\cite{fettiplace,mathai,shinoda}. 

To better understand the parameter dependence of the droplet velocity, we now perform a systematic parameter scan: as a result, we find that the droplet speed not only increases with contractility but also with the deformation resistance $D_{\phi}$ (Fig.~\ref{fig1}(c)), so that stiff circular cells move faster than easily deformable ones. 
Intriguingly, we also find a moderate friction with a substrate, $\Gamma>0$, increases the droplet speed whereas strong friction ($\Gamma\gg 1$) entirely suppresses motion as we shall see below. 

\begin{figure}
\includegraphics[width=0.485\textwidth]{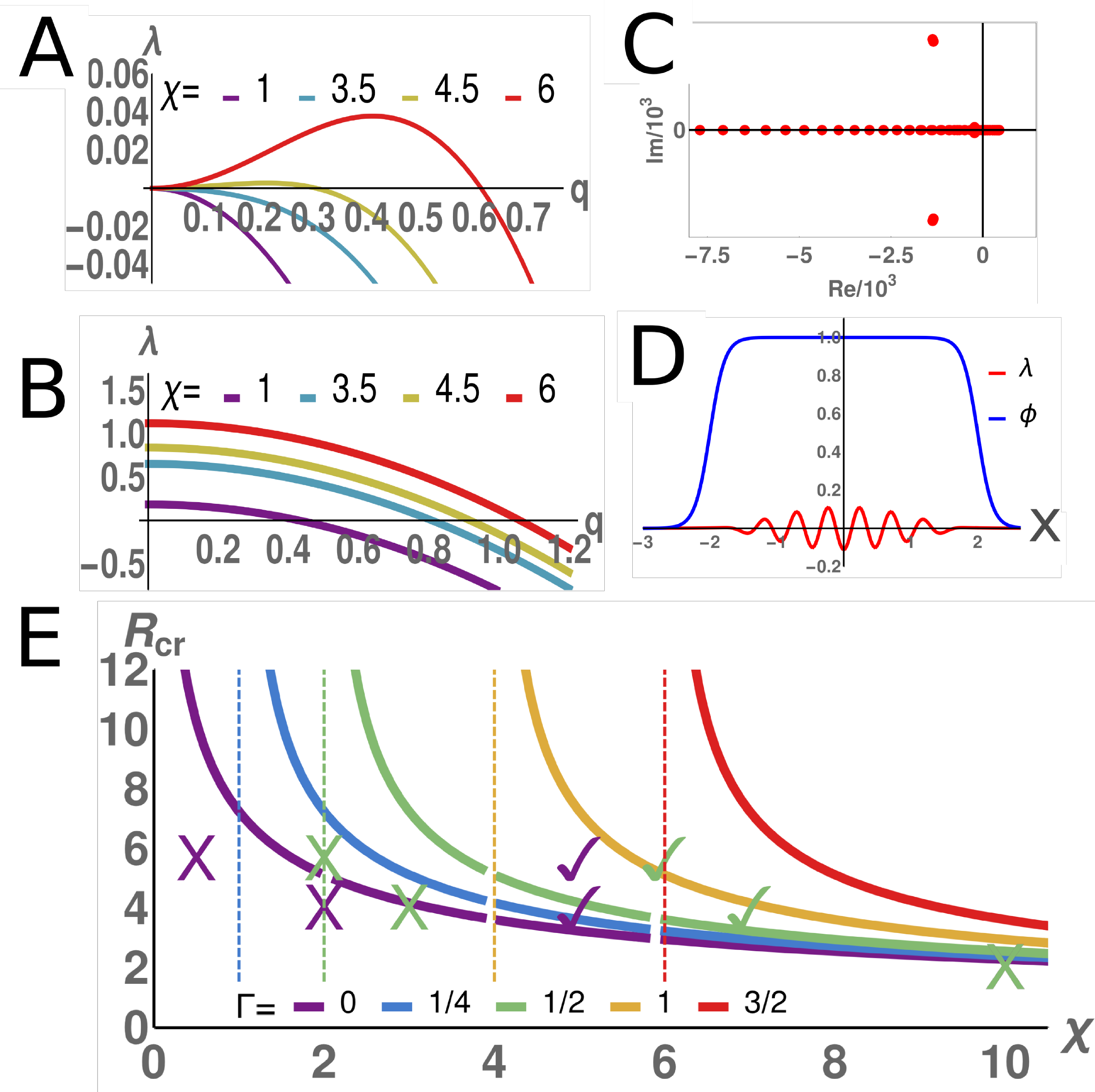}
\caption{Linear stability analysis: (A,B): $\phi=1$ (no droplet confinement);
growth rate $\lambda(q)$ of small fluctuations around the uniform phase with friction [$\Gamma=1$, (A)] and without [$\Gamma=0$, (B)]. 
(C,D): $\phi \neq 1$: Red dots show eigenvalues $\lambda_1..\lambda_N$ whose real parts determine the 
growth rate of actomyosin-fluctuations within a droplet. 
Dots right of the Re$=0$-line represent unstable modes; the fastest growing one is shown in panel (D) and represents instability within the droplet and its suppression at the droplet boundaries,
meaning that contractility-induced cell-motility can only occur in large enough cells (see SM \cite{SM} for details).
Parameters: $\chi=4.5; R=\sqrt{V/\pi}=2$ and $L=3; dx=0.01; N=601$ (for discretization); others as in simulations. 
Panel (E) shows the corresponding (phase) $\chi$-dependent critical cell radii for different friction values. 
Symbols $\checkmark$ (cell motion) and {\sffamily{X}} (no motion) show agreement with numerical simulations and dashed lines show critical $\chi$ values, 
below which contractility-induced motility is impossible, even for very large cells.}
\label{fig:linstab} 
\end{figure}

To understand the instability mechanism leading to contractility-induced motility, as well as the threshold value for $\chi$, we now perform a linear stability analysis. Considering an infinite system first, i.e. $\phi\equiv 1$, we find the following dispersion relation (Fig.~\ref{fig:linstab}a,b), describing the growth rate of small fluctuations around the uniform phase as a function of the wavenumber $q$~\cite{SM}:
\begin{equation}
\lambda({\bf q})={\bf q}^2 \left(\frac{\alpha_m\chi m_0}{(1+m_0)^2[\Gamma + (2+\eta){\bf q}^2]}-1 \right). \label{disp}
\end{equation}
Linear instability of the uniform phase occurs when (the real part of) $\lambda({\bf q})$ is positive for some wavevector ${\bf q}$ (in Fig.~\ref{fig:linstab}A, this corresponds to the red and the dark yellow curves), which leads to the instability criterion
\begin{equation}
\frac{\alpha_m\chi m_0}{\Gamma (1+m_0)^2}>1. \label{instabcrit}
\end{equation}
This result shows that the uniform phase is unstable to patterning if $\chi$ is strong enough (or simply if it is $>0$ in absence of friction, Fig.~\ref{fig:linstab}B). Increasing $m_0$ initially promotes the instability, but too large a value restores the uniform phase.

Extending our stability analysis to the case where a nonuniform phase field is present, we find that the above criterion still holds, but only if the cell is sufficiently large (see~\cite{SM} for details). In fact, as visualized in Fig.~\ref{fig:linstab}(D), the fastest growing mode (panel C), is localised in the center of the cell and gets suppressed at the boundaries. If the droplet is too small, the boundary suppression destroys myosin patterns, and the droplet is stationary. We quantify this argument by requiring that the shortest possible unstable wavelength (e.g. for the red line in panel A this is about $l\approx 2\pi/0.62$) is smaller that the diameter of the cell (2$R$) to allow for myosin accumulation within the droplet (and hence droplet motion). Through Eq.~\eqref{disp}, this leads to the critical contractility
\begin{equation}
\chi_c=\frac{(1+m_0)^2}{\alpha_mm_0}\left[\Gamma+(2+\eta)\left(\frac{\pi}{R}\right)^2\right]. 
\label{chic}
\end{equation}
For the parameters used in simulations presented in Fig.~\ref{fig1} ($V_{tar}=12.5$ and $\alpha_m=1$), the critical contractility is $\chi_c\leq13.2$ with $\Gamma=0$ and $\chi_c\leq14$ with $\Gamma=0.2$, in good agreement with our numerics. We show predictions from Eq.~(\ref{chic}) in  panel E and compare it with simulations. Eq.~\ref{chic} -- in dimensional units -- suggests that key control parameters are $\frac{{\mathcal{X}}}{\gamma D_m}$, for $\gamma\ne 0$, and $\frac{{\mathcal{X}} R^2}{\mu D_m}$, for $\gamma=0$: when these are large enough, the contractile isotropic droplet moves.

\begin{figure}
\centering
\includegraphics[width=6.cm]{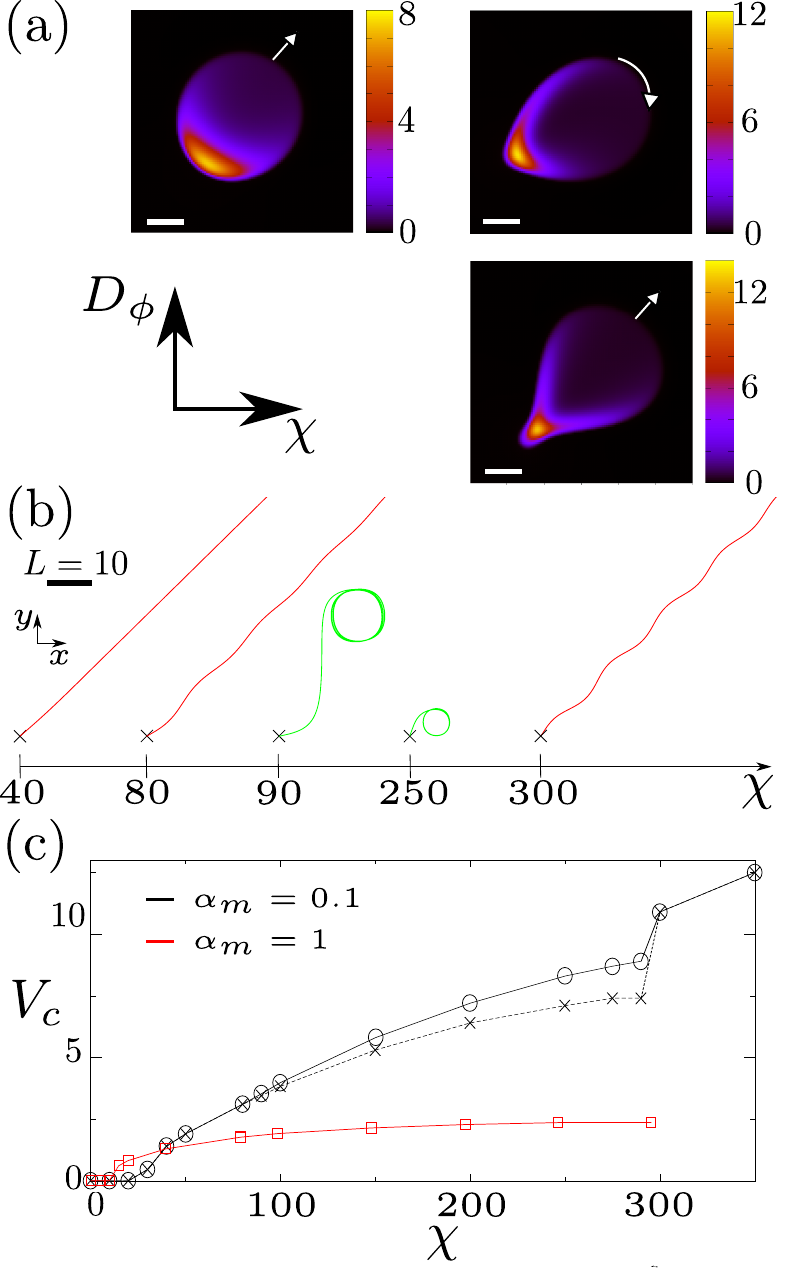}
\caption{Simulation results for $V_{tar}=12.5$, $\Gamma=0$ and $\alpha_{m}=0.1$. (a) Left top panel: myosin profile for $\chi=80$ and $D_{\phi}=20$. Right top panel: myosin profile for $\chi=200$ and $D_{\phi}=20$. Right bottom panel: myosin profile for $\chi=200$ and $D_{\phi}=5$. Scale bar is $1$. (b) Different trajectories of the droplet center depending on $\chi$ for $D_{\phi}=20$. (c) Velocity of the droplet as a function of contraction $\chi$ for $D_{\phi}=20$. The black solid line with circles corresponds to the droplet center of mass, the black dashed line with crosses to the myosin center of mass. The red curves with squares corresponds to $\alpha_m=1$.} 
\label{fig3}
\end{figure}

{\it Circular droplet motion} --
We now explore the case of low affinity between myosin and actin, $\alpha_m<1$, again for $\Gamma=0$. In this case, our droplets do not always swim straight, but may follow oscillatory trajectories or lock into a regular circular motion,  dependending on the value of $\chi$ and $D_{\phi}$ (see Fig.~3a).

What is the mechanism underlying deviations from a linear droplet motility? For $\alpha_m<1$ myosin is advected slower than the actin network, whose speed is approximately equal to the overall cell velocity. As a consequence of its slower speed, the myosin spot, which is elliptical for large $\alpha_m$ (Fig.~\ref{fig1}a,b), reshapes into a crescent-like form (see Fig.~3a), which is a consequence of myosin accumulation at the lateral cell boundaries. As we increase $\chi$, the crescent becomes longer and thinner. Crucially, for our noisy initial conditions the cresence 'grows' asymmetrically at both sides of the cell. This asymmetric growth results in a torque, since contraction takes place along the myosin concentration gradient, which pulls the cell perpendicular to its direction of motion, leading to curved motion. 
Remarkably, as the cell moves faster than myosin the curved motion further enhances the asymmetry of the crescent: thus, a sufficiently strong initial asymmetry of the crescent triggers a positive feedback loop between crescent asymmetry and cell-turning rate ultimately yielding circular motion.
This mechanism is only valid for a relatively undeformed cell. If $D_{\phi}$ is small, the cell can respond to the emerging torque simply by deforming, disrupting the feedback loop described above. This picture is in line with our simulations showing that for $\alpha_m<1$ and small $D_\phi$ the droplet forms a tail at the rear confining the myosin spot and hampering the formation of a large and asymmetric crescent (Fig.~\ref{fig3}a).

It is instructive to explore the droplet velocity $V_c$ as a function of $\chi$ for $\alpha_m<1$ (Fig.~\ref{fig3}(c)). For $\alpha_m=0.1$ the contractility threshold before cell motion sets in is much larger than for $\alpha_m=1$, as predicted by~(\eqref{chic}).
Interestingly, beyond this threshold, the reduced actin-myosin affinity leads to faster droplet motion. Finally, for strong contraction, when we reach the regime of circular motion, the velocity of the center of mass of the myosin cluster is smaller than $V_c$. This means that the myosin center of mass is closer to the middle of the trajectory than the cell center, consistent with our physical argument for circular motion. 

\begin{figure}
\centering
\includegraphics[width=9.cm]{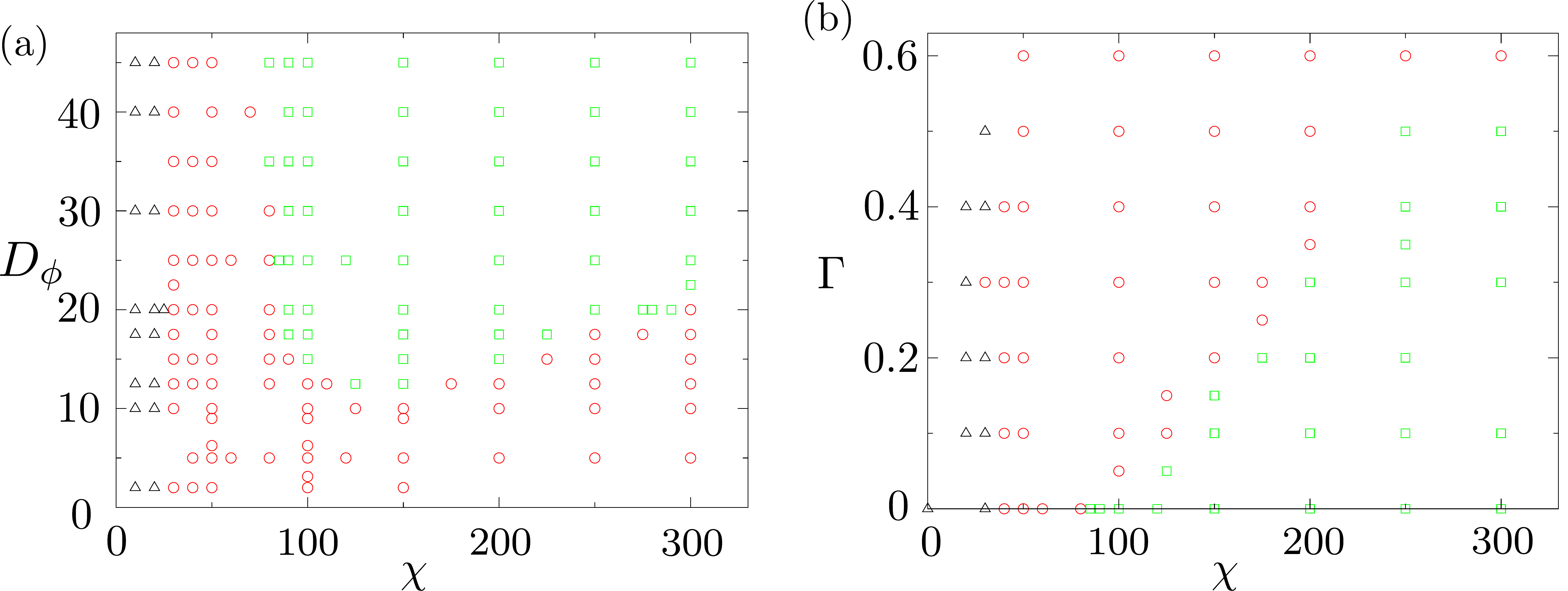}
\caption{(a) $\chi-D_{\phi}$ phase diagram for $\alpha_m=0.1$ and $\Gamma=0$. Black triangles, red circles and green squares respectively denote stationary cells, straight and curved motion. (b) $\chi-\Gamma$ phase diagram for $\alpha_m=0.1$ and $D_{\phi}=25$.} 
\label{fig4}
\end{figure}

To get a more comprehensive overview of the parameter regimes leading to straight, oscillatory or circular cell motion, we systematically performed a large number of simulations for different parameter regimes, and summarize our results in two phase diagrams, depending on  $\chi, D_{\phi}$ for in-bulk motility, $\Gamma=0$ (Fig.~\ref{fig4}(a)), and on $\chi, \Gamma$ for motion with friction ($\Gamma\ne 0$, Fig.~\ref{fig4}(b)). These diagrams show three different phases: (i) quiescent, (ii) rectilinear motion and (iii) circular motion. We find that friction favours rectilinear motion over circular one  in a similar way as large deformation resistance does. 


{\it Conclusions} --
In conclusion, we have proposed a generic mechanism exploiting motor-induced contractility to yield in-bulk motility of an isotropic actomyosin droplet. In-bulk motility arises when contractile stresses exceed a threshold scaling inversely with the cell size -- hence, even a very weak contractility may be enough to propel large droplets. While our mechanism is independent of the presence of a substrate, we have shown that wall-contact may both enhance the droplet speed or entirely suppress motion, subtly depending on parameters. 

Our in-bulk cell motility mechanism may apply to the motion of cells through 3D tissues {\it in vivo}, or through matrigel {\it in vitro}. It may also serve as a framework to design contractility-powered self-motile synthetic actomyosin droplets in the lab such as in~\cite{actomyosindroplet}.
Directly testable predictions of our work include the speed-up of motion with increasing contractility and stiffness, and the stabilization of oscillatory (circular) motion for large enough isotropic contractile stresses.



We thank EPSRC (grant EP/J007404/1) for support. BL gratefully acknowledges received funding by a Marie Sk{\l}odowska Curie Intra European Fellowship (G.A. no 654908) within Horizon 2020.




\end{document}


\title{Supplementary Material to: Actomyosin contraction induces droplet motility}
\author{T. Le Goff, B. Liebchen, D. Marenduzzo}
\date{\today}

\maketitle

Here, we discuss some details of the linear stability analysis which we use in the main text to predict
the transition between a non-motile phase (immobile droplet) and a motile one (moving droplet), both in absence (in bulk) and in the presence of substrate friction, as we will now see in detail:
\subsubsection{Uniform base state ($\phi=1$)}
We first identify the uniform solution $(m^\ast,{\bf v}^\ast)=(m_0,{\bf 0})$ of Eq. (3) in the main text which represents a nonmoving actomyosin field
with density $m_0$. Using the Ansatz $m=m^\ast+m',{\bf v}={\bf v}^\ast + {\bf v}'$ and 
linearizing the Eqs. (3) in the main text for $\phi=1$ around the uniform solution to understand the dynamics of small flucations, 
yields, after Fourier transforming the result: 
\begin{eqnarray}
\dot m' &=&m_0 \alpha_m {\rm i} ({\bf q} \cdot {\bf v}')-{\bf q}^2 m' \\
{\bf v}'&=&\frac{-{\rm i} {\bf q} \chi}{(1+m_0)^2 [\Gamma + (2+\eta){\bf q}^2]}
\end{eqnarray}
Combining these equations for the dynamics of the fluctuations $m',{\bf v}'$ and using the Ansatz $\dot m'({\bf q},t)={\rm exp}[\lambda(q) t] m'({\bf q},0)$ yields
the following dispersion relation for fluctuations around the uniform phase
\begin{equation}
\lambda({\bf q})={\bf q}^2 \left(\frac{\alpha_m \chi m_0}{(1+m_0)^2[\Gamma + (2+\eta){\bf q}^2]}-1 \right) \label{disp}
\end{equation}
Linear instability of the uniform solution occurs when (the real part of) $\lambda({\bf q})$ is positive for at least some wavevector ${\bf q}$, which 
leads to the instability criterion
\begin{equation}
\frac{\alpha_m \chi m_0}{\Gamma (1+m_0)^2}>1 \label{instabcrit}
\end{equation}
This criterion depenends only on $\chi \alpha_m,m_0$ and $\Gamma$.
If contraction is strong enough and enough myosin is present, with a large enough binding affinity (large $m_0,\chi,\alpha_m$), the positive feedback loop 
between myosin-induced 'fluid' advection and advection-induced myosin aggregation dominates substrate friction and the uniform phase looses stability in favour of actomyosin-aggregates. 
For $m_0=1,\alpha_m=1$, as used in most of our simulations, (\ref{instabcrit}) reduces to $\chi>4\Gamma$ (see Fig.~2 A, main text) suggesting the onset of cell motion at
$\chi>4$ for $\Gamma=1$. In absence of substrate friction ($\Gamma=0$), the myosin feedback loop has no competitor and any positive $\chi$ destabilizes the uniform phase. 
Note that in all cases, very large $m_0$ values suppress the instability; this represents the scenario where most actin fibres are saturated with myosin so that substantial deviations from 
a uniform myosin gradient are impossible. 
\\The fastest growing mode (maximum of $\lambda({\bf q})$), which determines the early-time length scale of structures (clusters) growing out of the 
uniform phase, results from (\ref{disp}) as 
\begin{equation}
q_{\rm max}=\left(\frac{\sqrt{\Gamma \frac{m_0 \chi}{(1+m_0)^2}}-\Gamma}{2+\eta}\right)^{1/2} \label{maxgr}
\end{equation}
The corresponding length scale $l_{\rm max}=2\pi/q_{\rm max}$ of contraction-induced structures increases with $\eta,\Gamma$ and 
decreases with $\chi$; that is, we expect large early-time structures close to the onset of instability and smaller ones further away from onset.

%

\subsubsection{Presence of a cell ($\phi \neq 1$)}
In the presence of droplet boundaries (cell membrane), $\phi$ builds up a nonuniform steady state profile given by the corresponding solution of Eq.~(3) in the main text.
Here, we calculate the growth rate of fluctuation around such a nonuniform state in one dimension and use 
the following approximate representation for the steady state solutions
\begin{equation}
m^\ast(x)=\phi^\ast(x)=\frac{1}{2}\left(1-\tanh \left[\sqrt{\frac{\Gamma}{8D_\phi}}(|x|-R)\right] \right)
\end{equation}
where $R$ is the radius of the cell, and $v^\ast=0$. Now we write $(m,\phi,v)=(m^\ast,\phi^\ast,v^\ast)+(m',\phi',v')$
and linearize the time-dependent equations of motion (Eqs. (3), main text) in the fluctuations $(m',\phi',v')$ around the nonuniform base-state $(m^\ast,\phi^\ast,v^\ast)$.
Representing the resulting equations on a grid $-L,-L+dx,...,L$, algebraically 
eliminating $v'$ and using the Ansatz $m'_i(t)={\rm exp}(\lambda_i t)m'_i(0),\phi'_i(t)={\rm exp}(\lambda_i t)\phi'_i(0)$
yields a $N=2L/dx+1$-dimensional matrix-vector equation which we solve for the eigenvalues $\lambda_1,..\lambda_N$ by numerical diagonalization. 
We visualize the result of this procedure in Fig.~2 C,D (main text) for $\chi=4.5$ (i.e. close to the onset 
of instability in the corresponding uniform system). Here, panel C shows that a few of the eigenvalues have a positive 
real part, i.e. the contraction-induced linear instability survives the presence of droplet boundaries and leads to 
a narrow band of unstable modes close to $\chi=4$. Panel D visualizes the mode with the largest growth rate in configuration 
space (red) alongside the base phase field (blue). 
Here, deep inside the cell, the wavelength of the shown mode resembles the one of the fastest growing mode of the 
underlying uniform system (\ref{maxgr}).
However, the figure also shows that instability exists only in the interior of the cell where the actomyosin concentration is highest but is
suppressed at the cell-boundaries. (Note that when the cell starts to deform (or move), the maximum of the actomyosin concentration may leave the cell center and the instability might be most effective close to the
cell boundaries.)
\\This finding of suppression of instability close to the cell-boundaries suggests that instability is entirely suppressed 
if the cell is too small, i.e. the present linear stability analysis suggests that small cells cannot move based on myosin-contraction. We therefore ask: 
What is the critical cell size to obtain instability and contraction-induced droplet-motility?
We first note that instability can only occur if the shortest unstable mode 
(of the instability band of the underlying uniform system) is smaller than the droplet size. 
Thus, we predict the critical cell-size 
as $l=2\pi/q_c$ where $q_c$ is the short wavelength edge of the instability band of the underlying uniform system
(i.e. the point where $\lambda(q)$ crosses the $\lambda=0$ axis in Fig.~2 A,B, main text).
We can readily calculate $q_c$ from the dispersion relation (\ref{disp}) and obtain the critical cell radius 
$R_{\rm cr}$ from the condition that at least one wavelength of the shortest possible unstable mode fits into the cell, i.e. from
$2R_{\rm cr}=2\pi/q_c$\footnote{We note that a complete numerical linear stability in presence of the phase field suggests that in many cases also half a wavelength 
can build up in the cell, suggesting an somewhat 'earlier' onset of cell-motility than predicted below in terms of $R_{\rm cr}$.}, as 
\begin{equation}
R_{\rm cr}=\pi\sqrt{\frac{2+\eta}{\frac{\chi m_0}{(1+m_0)^2}-\Gamma}} 
\end{equation}
We visualize this critical cell size in an instability diagram (or nonequilibrium phase diagram) in Fig.~2E (main text) and find very good agreement with direct 
numerical simulations of the equations of motion. Our simulations also confirm the predicted 
$R_{\rm cr}\propto 1/\sqrt{\chi}$
scaling.